%% file: LUP-INAP-11.tex
\documentclass{llncs}
\usepackage{a4,a4wide}
\usepackage{amsmath,amssymb}
\usepackage{relsize}
\usepackage{paralist}
\usepackage{graphicx}
\usepackage{xspace}
\usepackage{algorithm}
\usepackage{algorithmic}
\usepackage{times} 
\usepackage{helvet} 
\usepackage{courier}
\usepackage{subfig}

\input{mx-def.tex}

\title{Lifted Unit Propagation for Effective Grounding} 
\author{Pashootan Vaezipoor\inst{1} \and David Mitchell\inst{1} \and Maarten Mari\"en\thanks{This author's contributions to this paper were made while he was a post-doctoral fellow
 at SFU.}\inst{2}} 
\institute{Department of Computing Science, Simon Fraser University, Canada \email{\{pva6,mitchell\}@cs.sfu.ca} \and Department of Computer Science, Katholieke Universiteit Leuven, Belgium \email{maartenm@cs.kuleuven.be}}

\begin{document} 

\maketitle 
\thispagestyle{empty}
\pagestyle{empty}

\begin{abstract}
A grounding of a formula $\phi$ over a given finite domain is 
a ground formula which is equivalent to $\phi$ on that domain.
Very effective propositional solvers have made grounding-based
methods for problem solving increasingly important, however for 
realistic problem domains and instances, the size of groundings 
is often problematic.  A key technique in ground (e.g., SAT) solvers 
is unit propagation, which often significantly reduces ground 
formula size even before search begins.  
We define a ``lifted'' version of unit propagation which may be carried 
out prior to grounding, and describe integration of the 
resulting technique into grounding algorithms. 
We describe an implementation of the method in a bottom-up 
grounder, and an experimental study of its performance.
\end{abstract}

\section{Introduction\label{sec:Intro}}
Grounding is central in many systems for solving combinatorial problems based on declarative specifications.   In grounding-based systems, a ``grounder'' combines a 
problem specification with a problem instance to produce a ground formula which represents the solutions for the instance.  A solution (if there is one) is obtained 
by sending this formula to a ``ground solver'', such as a SAT solver or 
propositional answer set programming (ASP) solver.
Many systems have specifications given in extensions or restrictions of classical 
first order logic (FO), including: IDP \cite{IDP}, MXG \cite{Mohebali:MSc-2006}, 
Enfragmo \cite{DBLP:conf/lpar/AavaniTUTM10,DBLP:conf/ai/AavaniWTM11}, ASPPS \cite{EastTruszczynski:2006}, and Kodkod \cite{DBLP:conf/tacas/TorlakJ07}.
Specifications for ASP systems, such as DLV \cite{LeonePFEGPS06} and clingo \cite{gekakaosscth08b}, 
are (extended) normal logic programs under stable model semantics.   

Here our focus is grounding specifications in the form of FO formulas. 
In this setting, formula $\phi$ constitutes a specification of a problem 
(e.g., graph 3-colouring), and a problem instance is a finite structure \cA (e.g., a graph).  The grounder, roughly, must produce a ground formula $\psi$ which 
is logically equivalent to $\phi$ over the domain of \cA.  
Then $\psi$ can be transformed into a propositional CNF formula, and given 
as input to a SAT solver.  If a satisfying assignment is found, a solution to 
\cA can be constructed from it.   ASP systems use an analogous process.

A ``naive'' grounding of $\phi$ over a finite domain $A$ can be obtained by 
replacing each sub-formula of the form 
$\exists x\, \psi(x)$ with $\bigvee_{a\in A} \psi(\tilde{a})$, where $\tilde{a}$ is a
constant symbol which denotes domain element $a$, and similarly replacing 
each subformula $\forall x\, \psi(x)$ with a conjunction.
For a fixed FO formula $\phi$, this can be done in time polynomial in $|A|$.   
Most grounders use refinements of this method, implemented top-down or 
bottom-up, and perform well on simple benchmark problems and small instances.  
However, as we tackle more realistic problems with complex specifications and 
instances having large domains, the groundings produced can become prohibitively 
large.   This can be the case even when the formulas are ``not too hard''.
That is, the system performance is poor because of time spent generating and manipulating this large ground formula, yet an essentially equivalent but 
smaller formula can be solved in reasonable time.   
This work represents one direction in our group's efforts to develop techniques which 
scale effectively to complex specifications and large instances.

Most SAT solvers begin by executing unit propagation (UP) on the input formula
(perhaps with other ``pre-processing'').  This initial application 
of UP often eliminates a large number of variables and clauses, 
and is done very fast.   However, it may be too late: the system has already 
spent a good deal of time generating large but rather uninteresting (parts of) 
ground formulas, transforming them to CNF, moving them from the grounder to 
the SAT solver, building the SAT solver's data structures, etc.
This suggests trying to execute a process similar to UP  
{\em before or during grounding}.  

One version of this idea was introduced in \cite{WittocxMD:08,DBLP:journals/jair/WittocxMD10}.
The method presented there involves computing 
a {\em symbolic} and {\em incomplete} representation of the information that 
UP could derive, obtained from $\phi$ alone without reference to a particular 
instance structure.   For brevity, we refer to that method as GWB, for 
``Grounding with Bounds''.  In \cite{WittocxMD:08,DBLP:journals/jair/WittocxMD10}, 
the top-down grounder \gidl \cite{197022} is 
modified to use this information, and experiments indicate it significantly reduces 
the size of groundings without taking unreasonable time.

An alternate approach is to construct a {\em concrete} and {\em complete} 
representation of the information that UP can derive about a grounding of 
$\phi$ over \cA, and use this information during grounding to reduce grounding 
size.   This paper presents such a method, which we call {\em lifted unit propagation} (LUP).  
(The authors of the GWB papers considered this approach also \cite{PersonalCommunication2008},
but to our knowledge did not implement it or report on it.
The relationship between  GWB and LUP is discussed further in Section \ref{sec:disc}.)
The LUP method is roughly as follows.  
\begin{enumerate}
\item Modify instance structure \cA to produce a new (partial) 
 structure which contains information equivalent to that derived by 
 executing UP on the CNF formula obtained from a grounding of $\phi$ over \cA.
 We call this new partial structure the LUP structure for $\phi$ and \cA, 
 denoted ${\cal LUP}$($\phi$,\cA).
\item Run a modified (top-down or bottom-up) grounding algorithm which takes as 
 input, $\phi$ and ${\cal LUP}$($\phi$,\cA), and produces a grounding of 
 $\phi$ over \cA.   
\end{enumerate}

The modification in step 2 relies on the idea that a tuple in ${\cal LUP}$($\phi$,\cA) indicates that a particular subformula has the same (known) truth value in every model. Thus, that subformula may be replaced with its truth value. The CNF formula obtained by grounding over ${\cal LUP}$($\phi$,\cA) is at 
most as large as the formula that results from producing the naive grounding 
and then executing UP on it.   Sometimes it is much smaller than this, because 
the grounding method naturally eliminates some autark sub-formulas which 
UP does not eliminate, as explained in Sections~\ref{sc:tdgnd} and~\ref{sec:exp}.
 
We compute the LUP structure by constructing, from $\phi$, an inductive 
definition of the relations of the LUP structure for $\phi$ and \cA (see Section \ref{sec:LUP}).  We 
implemented a semi-naive method for evaluating this inductive definition, 
based on relational algebra, within our grounder Enfragmo.  
(We also computed these definitions using the 
ASP grounders gringo and DLV, but these were not faster. )

For top-down grounding (see Section \ref{sc:tdgnd}), we modify the naive recursive algorithm to check 
the derived information in ${\cal LUP(\phi,\cA)}$ at the time of instantiating 
each sub-formula of $\phi$.   This algorithm is presented primarily for 
expository purposes, and is similar to the modified top-down algorithm 
used for GWB in \gidl.

For bottom-up grounding (see Section \ref{sc:bugnd}), we revise the bottom-up grounding 
method based on extended relational algebra described in \cite{U-SFraser-CMPT-TR:2006-24,PLTG:2007:IJCAI}, which is the basis of 
grounders our group has been developing. The change 
required to ground using ${\cal LUP(\phi,\cA)}$ is a simple revision to the base case.

In Section \ref{sec:exp} we present an experimental evaluation of the performance of our grounder 
Enfragmo with LUP.  This evaluation is limited by the fact that our LUP 
implementation does not support specifications with arithmetic or aggregates,
and a shortage of interesting benchmarks which have natural specifications 
without these features.   Within the limited domains we have tested to date, we found:

\begin{compactenum}
\item 
CNF formulas produced by Enfragmo with LUP are always 
smaller than the result of running UP on the CNF formula produced 
by Enfragmo without LUP, and in some cases much smaller.
\item  
CNF formulas produced by Enfragmo with LUP are always 
smaller than the ground formulas produced by \gidl\!, with or without 
GWB turned on.
\item 
Grounding over ${\cal LUP(\phi,\cA)}$ is always slower than grounding 
without, but CNF transformation with LUP is almost always faster than 
without.
\item 
Total solving time for Enfragmo with LUP is sometimes significantly 
less than that of Enfragmo without LUP, but in other cases is somewhat greater.
\item 
Enfragmo with LUP and the SAT solver MiniSat always 
runs faster than the IDP system (\gidl with ground solver \textsc{MiniSat(ID)}), 
with or without the GWB method turned on in \gidl.
\end{compactenum}
Determining the extent to which these observations generalize is future work.

\section{FO Model Expansion and Grounding \label{sec:Background}}

A natural formalization of combinatorial search problems and their specifications 
is as the logical task of model expansion (MX) \cite{Mitchell2011}. 
Here, we define MX for the special case of FO.
Recall that a structure \cB for vocabulary $\sigma \cup \varepsilon$ is an 
expansion of $\sigma$-structure \cA 
 iff  
\cA and \cB have the same domain ($A = B$), and interpret their common 
vocabulary identically, i.e., for each symbol $R$ of $\sigma$, $R^{\cB} = R^{\cA}$. 
Also, if \cB is an expansion of $\sigma$-structure \cA, then \cA is the reduct 
of \cB defined by $\sigma$.

\begin{definition}[Model Expansion for FO]
\begin{description}
\item \underline{Given:} A FO formula $\phi$ on vocabulary $\instvocab \cup \expvocab$ and a $\sigma$-structure \cA,
\item \underline{Find:} an expansion $\model$ of $\given$ that satisfies $\phi$.
\end{description}
\end{definition}

In the present context, 
the formula $\phi$ constitutes a problem specification, the structure \cA a 
problem instance, and expansions of \cA which satisfy $\phi$ are solutions 
for \cA.  Thus, we call the vocabulary of \cA, the \emph{instance} 
vocabulary, denoted by $\sigma$, and 
$\expvocab$ the \emph{expansion} vocabulary. We sometimes say $\phi$ is $\given$-satisfiable if there exists an 
expansion \cB  of \cA that satisfies $\phi$. 

\begin{example}
Consider the following formula $\phi$:
\begin{equation*}
\forall x [(R(x) \! \vee \! B(x)\! \vee \! G(x))
\wedge \neg(R(x)\! \wedge \! B(x)) \wedge \neg(R(x)\! \wedge \! G(x))\wedge 
\neg(B(x)\! \wedge \! G(x))]
\end{equation*}
\vspace{-5mm}
\begin{equation*}
\wedge \ \ \forall  x\forall y [E(x, y)\supset (\neg(R(x)\! \wedge \! R(y))
\wedge \neg (B(x)\! \wedge \! B(y)) \wedge \neg (G(x) \! \wedge \! G(y)))].
\end{equation*}

A finite structure \cA over vocabulary $\instvocab = \{E\}$, where $E$ is a binary 
relation symbol, is a graph.
Given graph $\given= {\cal{G}} = (V;E)$, there is an expansion $\model$ of $\given$ that satisfies $\phi$, iff $\cal{G}$ is 3-colourable. 
So $\phi$ constitutes a specification of the problem of graph 3-colouring. To illustrate:
\[
\underbrace{\overbrace{(V; E^{\given}}^{\given}, R^{\model}, B^{\model}, G^{\model})}_{\model} \models \phi
\]

\noindent An interpretation for the expansion vocabulary $\expvocab := \{R, B, G\}$ given by structure $\model$ is a colouring of $\cal G$, and the proper 3-colourings of $\cal G$ 
are the interpretations of $\expvocab$ in structures $\model$ that satisfy $\phi$.
\end{example}

\subsection{Grounding for Model Expansion}

Given $\phi$ and \cA, we want to produce a CNF formula (for input to a SAT 
solver), which represents the solutions to \cA.   We do this in two steps: grounding, followed by transformation to CNF. The grounding step produces a ground formula $\psi$ which is equivalent 
to $\phi$ over expansions of \cA.
To produce $\psi$, we bring domain elements into the syntax by expanding the vocabulary with a new constant symbol for each domain element.  
For $\givendom$, the domain of $\given$, we denote this set of constants by 
$\tilde{\givendom}$.  
For each $a \in A$, we write $\tilde{a}$ for the corresponding symbol in $\tilde{A}$.
We also write $\tilde{\bar{a}}$, where $\bar{a}$ is a tuple.

\begin{definition}[Grounding of $\phi$ over \cA]
\label{def:grounding}
Let $\phi$ be a formula of vocabulary $\sigma \cup \varepsilon$, 
\cA be a finite $\sigma$-structure, and $\psi$ be a ground formula 
of vocabulary $\mu$, where $\mu \supseteq \sigma \cup \varepsilon \cup \tilde{A}$. 
Then $\psi$ is a grounding of $\phi$ over \cA if and only if:
\begin{compactenum}
\item if $\phi$ is \cA-satisfiable then $\psi$ is \cA-satisfiable;
\item if $\cB$ is a $\mu$-structure which is an expansion of \cA and gives 
$\tilde{A}$ the intended interpretation, 
  and $\cB \models \psi$, then $\cB \models \phi$.
\end{compactenum}
We call $\psi$ a reduced grounding if it contains no symbols of the instance 
vocabulary $\sigma$.
\end{definition}

Definition~\ref{def:grounding} is a slight generalization of that used  
in \cite{U-SFraser-CMPT-TR:2006-24,PLTG:2007:IJCAI}, in that it allows $\psi$ to 
have vocabulary symbols not in $\sigma \cup \varepsilon \cup \tilde{A}$.  
This generalization allows us to apply a 
Tseitin-style CNF transformation in such a way that the resulting 
CNF formula is still a grounding of $\phi$ over \cA.
If $\model$ is an expansion of $\given$ satisfying $\psi$, then the reduct of $\model$ 
defined by $\sigma \cup \varepsilon$ is an expansion of $\given$ that satisfies $\phi$.  
For the remainder of the paper, we assume that $\phi$ is in negation 
normal form (NNF), i.e.,  negations are applied only to atoms.
Any formula may be transformed in linear time to an equivalent formula 
in NNF.

Algorithm~\ref{alg:top-down-naive} produces the ``naive grounding'' of $\phi$ 
over \cA mentioned in the introduction. A substitution is 
a set of pairs $(x/a)$, where $x$ is a variable and $a$ a constant symbol.
If $\theta$ is a substitution, then $\phi[\theta]$ denotes the result of 
substituting constant symbol $a$ for 
each free occurrence of variable $x$ in $\phi$, for every $(x/a)$ in $\theta$.   
We allow conjunction and disjunction to be connectives of 
arbitrary arity.   That is $(\land\ \phi_1\ \phi_2\ \phi_3)$
is a formula, not just an abbreviation for some parenthesization of 
$(\phi_1 \land \phi_2 \land \phi_3)$. The initial call to Algorithm \ref{alg:top-down-naive} is 
$\mathrm{NaiveGnd}_{\cA}(\phi,\emptyset)$, 
where $\emptyset$ is the empty substitution.

\begin{algorithm}
\caption{Top-Down Naive Grounding of NNF formula $\phi$ over $\given$}
\label{alg:top-down-naive}
\[
\mathrm{NaiveGnd}_{\cA}(\phi,\theta)\!=\!\!\begin{cases}
P(\bar{x})[\theta]  & \mathrm{if }\ \phi \mbox{ is an atom }  P(\bar{x}) \\
\neg P(\bar{x})[\theta]  & \mathrm{if }\  \phi \mbox{ is a negated atom } \neg P(\bar{x}) \\
\bigwedge_i \mathrm{NaiveGnd}_{\cA}(\psi_{i},\theta) & \mathrm{if }\  \phi=\bigwedge_i \psi_i \\
\bigvee_i \mathrm{NaiveGnd}_{\cA}(\psi_{i},\theta) & \mathrm{if }\  \phi=\bigvee_i \psi_i \\
\bigwedge_{a \in A} 
   \mathrm{NaiveGnd}_{\cA}(\psi,[\theta \cup (x/\tilde{a})]) & \mathrm{if }\ \phi=\forall x\ \psi \\
\bigvee_{a \in A} 
   \mathrm{NaiveGnd}_{\cA}(\psi,[\theta \cup (x/\tilde{a})]) & \mathrm{if }\ \phi=\exists x \ \psi
\end{cases}
\]
\end{algorithm}

The ground formula produced by Algorithm~\ref{alg:top-down-naive} is not 
a grounding of $\phi$ over \cA (according to Definition \ref{def:grounding}), because 
it does not take into account the interpretations of $\instvocab$ 
given by \cA.   To produce a grounding of $\phi$ over \cA, we may conjoin 
a set of atoms giving that information.   In the remainder of the paper, 
we write $\mathrm{NaiveGnd}_{\cA}(\phi)$ for the result of calling 
$\mathrm{NaiveGnd}_{\cA}(\phi,\emptyset)$ and conjoining ground atoms to 
it to produce a grounding of $\phi$ over \cA.
We may also produce a reduced grounding from 
$\mathrm{NaiveGnd}_{\cA}(\phi,\emptyset)$ by ``evaluating out'' all 
atoms of the instance vocabulary. The groundings produced by algorithms described later in this paper can be obtained by simplifying 
out certain sub-formulas of $\mathrm{NaiveGnd}_{\cA}(\phi)$.

\ignore{
\[
Gnd(\phi(\bar{x}), \bar{a})\!=\!\!\begin{cases}
P(\bar{a}) & \phi=P(\bar{x}) \\
\neg P(\bar{a}) & \phi=\neg P(\bar{x}) \\
\bigwedge_i Gnd(\psi_{i}(\bar{x}_{i}), \bar{a}|_{\bar{x}_{i}}) & \phi=\bigwedge_i \psi_i(\bar{x}_{i}) \\
\bigvee_i Gnd(\psi_{i}(\bar{x}_{i}), \bar{a}|_{\bar{x}_{i}}) & \phi=\bigvee_i \psi_i(\bar{x}_{i}) \\
\bigwedge_{a'\in \givendom}  Gnd(\psi(\bar{x},y),\bar{a}\cup a') &\phi=\forall y\ \psi(\bar{x},y) \\
\bigvee_{a'\in \givendom} Gnd(\psi(\bar{x},y),\bar{a}\cup a')& \phi=\exists y \ \psi(\bar{x},y)
\end{cases}
\]
\noindent Here, $\bar{a}$ is a tuple of domain elements. 
Abusing the notation, we use $\bar{a}\cup a'$ to 
refer to the tuple that is created by adding the new element $a'$ to it, 
and we use $\bar{a}|_{\bar{x}}$ to
restrict the elements in $\bar{a}$ to those that are referred to by variables in $\bar{x}$. 
} 

\subsection{Transformation to CNF and Unit Propagation}
\label{sec:CNF-Transform}

To transform a ground formula to CNF, we employ the method of Tseitin \cite{Tseitin} with two  modifications.   The method, usually presented for propositional formulas, 
involves adding a new atom corresponding to each sub-formula.  
Here, we use a version for ground FO formulas, so the resulting CNF formula 
is also a ground FO formula, over vocabulary 
$\tau = \sigma \cup \varepsilon \cup \tilde{A} \cup \omega$, where 
$\omega$ is a set of new relation symbols which we call ``Tseitin symbols''.  
To be precise, $\omega$ consists of a new $k$-ary relation symbol $\subfor{\psi}$ for each subformula $\psi$ of $\phi$ with $k$ free variables.
We also formulate the transformation for formulas in which conjunction and 
disjunction may have arbitrary arity.  
  
Let $\gamma = \mathrm{NaiveGnd}_{\cA}(\phi,\emptyset)$.
Each subformula $\alpha$ of $\gamma$ is a grounding over \cA of 
a substitution instance $\psi(\bar{x})[\theta]$, of some 
subformula $\psi$ of $\phi$ with free variables $\bar{x}$.
To describe the CNF transformation, it is useful to think of labelling the 
subformulas of $\gamma$ during grounding as follows. 
If $\alpha$ is a grounding of formula $\psi(\bar{x})[\theta]$, 
label $\alpha$ with the ground atom $\ceil{\psi}(\bar{x})[\theta]$.
To minimize notation, we will denote this atom by $\widehat{\alpha}$, setting $\widehat{\alpha}$ to $\alpha$ if $\alpha$ is an atom.
Now, we have for each sub-formula $\alpha$ of the ground formula $\psi$, a unique 
ground atom $\widehat{\alpha}$, and we carry out the Tseitin transformation to CNF using
these atoms.

\begin{definition}
For ground formula $\psi$, we denote by $\mathrm{CNF(\psi)}$ the following set of ground 
clauses.   For each sub-formula $\alpha$ of $\psi$ of form 
$(\land_i \ \alpha_i$), include in $\mathrm{CNF(\psi)}$ the set of clauses 
$\{ (\neg \widehat{\alpha} \vee \widehat{\alpha_i} ) \} \cup \{ (\lor_i \neg \widehat{\alpha_i} \vee \widehat{\alpha}) \}$, 
and similarly for the other connectives. 


\end{definition}

If $\psi$ is a grounding of $\phi$ over \cA, then 
CNF$(\psi)$ is also. The models of $\psi$ are exactly the reducts of the models of 
CNF$(\psi)$ defined by $\sigma \cup \expvocab \cup \tilde{A}$. CNF$(\psi)$ can trivially be viewed as a propositional CNF formula. This propositional formula can be sent to a SAT solver, 
and if a satisfying assignment is found, a model of $\phi$ which is an expansion of \cA
can be constructed from it.

\begin{definition}[UP$(\gamma)$]
Let $\gamma$ be a ground FO formula in CNF. Define $\mathrm{UP(\gamma)}$, the result of applying unit propagation to $\gamma$, 
to be the fixed point of the following operation: 

\begin{quote}
If $\gamma$ contains a unit clause $(l)$, delete from each clause of $\gamma$ every 
occurrence of $\neg l$, and delete from $\gamma$ every clause containing $l$.
\end{quote}
\end{definition}

Now, $\mathrm{CNF(NaiveGND_{\cA}(\phi))}$ is the result of producing the naive grounding of $\phi$ over \cA, and transforming it to CNF in the standard way, 
and $\mathrm{UP(\mathrm{CNF}(\mathrm{NaiveGND}_{\cA}(\phi)))}$ is the formula obtained after simplifying it 
by executing unit propagation.    These two formulas provide reference points for measuring the reduction in ground formula size 
obtained by LUP. 

\section{Bound Structures and Top-down Grounding}\label{sc:tdgnd}

We present grounding algorithms, in this section and in 
Section~\ref{sec:LUP}, which produce groundings of 
$\phi$ over a class of partial structures, which we call bound structures, 
related to \cA. The structure ${\cal LUP}$($\phi$,\cA) is a particular bound structure.   
In this section, we define partial structures and bound structures,
and then present a top-down grounding algorithm.  The formalization 
of bound structures here, and of ${\cal LUP}$($\phi$,\cA) in 
Section~\ref{sec:LUP}, are ours, although a similar formalization was implicit 
in \cite{PersonalCommunication2008}.

\subsection{Partial Structures and Bound Structures}

A relational $\tau$-structure \cA consists of a domain $A$ together with 
a relation $R^{\cA}\! \subset \!A^k$ for each $k$-ary relation symbol of $\tau$.
To talk about partial structures, in which the interpretation of a relation symbol 
may be only partially defined, it is convenient to view a structure in terms 
of the characteristic functions of the relations.   Partial $\tau$-structure \cA consists of a domain $A$ together with a $k$-ary function 
$\mathlarger{\chi}_{R}^{\cA} : A^{k}\to \{\top, \bot, \infty\}$, for each $k$-ary 
relation symbol $R$ of $\tau$.   Here, as elsewhere, $\top$ denotes true, 
$\bot$ denotes false, and $\infty$ denotes undefined.   If each of these 
characteristic functions is total, then \cA is total.   We may 
sometimes abuse terminology and call a relation partial, meaning 
the characteristic function interpreting the relation symbol in question is partial.

Assume the natural adaptation of standard FO semantics the to the case of partial relations, e.g. with Kleene's 3-valued semantics \cite{Kleene52}. For any (total) $\tau$-structure \cB, each $\tau$-sentence $\phi$ is either true 
or false in \cB ($\cB \models \phi$ or $\cB \not\models \phi$), and each 
$\tau$-formula $\phi(\bar{x})$ with free variables $\bar{x}$, defines a relation 
\begin{equation}\label{eq:relonA}
 \phi^{\model} = \{ \bar{a} \in \givendom^{|\bar{x}|} : \model \models \phi(\bar{x}) [\bar{x}/\bar{a}]\}.
 \end{equation}
Similarly, for any partial $\tau$-structure, each $\tau$-sentence is either 
true, false or undetermined in $\cB$, and each $\tau$-formula 
$\phi(\bar{x})$ with free variables $\bar{x}$ defines a partial function
\begin{equation}\label{eq:charfunc}
\mathlarger{\chi}_{\phi}^{\given}: A^{k}\to \{\top, \bot, \infty\}.
\end{equation}
In the case $\mathlarger{\chi}_{\phi}^{\given}$ is total, it is the characteristic function of 
the relation~(\ref{eq:relonA}).

There is a natural partial order on partial structures for any vocabulary 
$\tau$, which we may denote by $\leq$, where $\cA \leq \cB$ iff 
$\cA$ and $\cB$ agree at all points where they are both defined, and 
$\cB$ is defined at every point $\cA$ is.    If $\cA \leq \cB$, we may 
say that \cB is a strengthening of \cA.    When convenient, if the 
vocabulary of \cA is a proper subset of that of \cB, we may still call 
\cB a strengthening of \cA, taking \cA to leave all symbols not in 
its vocabulary, completely undefined.   
We will call \cB a conservative strengthening of \cA with respect to
formula $\phi$ if \cB is a strengthening of \cA and in addition every 
total structure which is a strengthening of \cA and a model of $\phi$ 
is also a strengthening of \cB.  (Intuitively, we could ground $\phi$ over 
\cB instead of \cA, and not lose any intended models.)

The specific structures of interest are over a vocabulary expanding the vocabulary 
of $\phi$ in a certain way.   We will call a vocabulary $\tau$ a Tseitin vocabulary 
for $\phi$ if it contains, in addition to the symbols of $\phi$, the set 
$\omega$ of Tseitin symbols for $\phi$.    We call a $\tau$-structure a 
``Tseitin structure for $\phi$'' if the interpretations of the Tseitin symbols 
respect the special role of those symbols in the Tseitin transformation. 
For example, if $\alpha$ is $\alpha_{1} \wedge \alpha_{2}$, then $\widehat{\alpha}^{\given}$
must be true iff $\widehat{\alpha_{1}}^{\given} = \widehat{\alpha_{2}}^{\given} = true$.
The vocabulary 
of the formula $\mathrm{CNF(NaiveGnd_{\cA}(\phi))}$ is a Tseitin vocabulary 
for $\phi$, and every model of that formula is a Tseitin structure for $\phi$.

\begin{definition}[Bound Structures]\label{def:bound-struct}
Let $\phi$ be a formula, and \cA be a structure for a sub-set of the 
vocabulary of $\phi$. 
A bound structure for $\phi$ and \cA is a partial  
Tseitin structure for $\phi$ that is a conservative strengthening of \cA 
with respect to $\phi$.
\end{definition}

Intuitively, a bound structure provides a way to represent the information 
from the instance together with additional information, including information 
about the Tseitin symbols in a grounding of $\phi$, that we may derive 
(by any means), provided that information does not eliminate any intended 
models.

Let $\tau$ be the minimum vocabulary for bound structures for $\phi$ and \cA.
The bound structures for $\phi$ and \cA with vocabulary $\tau$ form a lattice 
under the partial order $\leq$, with \cA the minimum element.   The maximum 
element is defined exactly for the atoms of $\mathrm{CNF(NaiveGnd_{\cA}(\phi))}$
which have the same truth value in every Tseitin $\tau$-structure that satisfies $\phi$. This is the structure produced by ``Most Optimum Propagator'' in \cite{DBLP:journals/jair/WittocxMD10}).

\begin{definition}[Grounding over a bound structure]\label{def:gndpar}
Let $\hat{\given}$ be a bound structure for $\phi$ and \cA.
A formula $\psi$, over a Tseitin vocabulary for $\phi$ which includes 
$\tilde{A}$, is a grounding of 
$\phi$ over $\hat{\given}$ iff 
\begin{compactenum}
\item if there is a total strengthening of $\hat{\given}$ that satisfies $\phi$,
 then there is a one that satisfies $\psi$;
\item if \cB is a total Tseitin structure for $\phi$ which strengthens $\hat{\given}$, gives $\tilde{A}$ the intended 
interpretation and satisfies $\psi$, then it satisfies $\phi$.
\end{compactenum}
\end{definition}

A grounding $\psi$ of $\phi$ over $\hat{\given}$ need not be a grounding of $\phi$ over  $\given$. If we conjoin with $\psi$ ground atoms representing the information contained in $\hat{\given}$, then we do obtain a grounding of $\phi$ over $\given$ . In practice, we send just CNF$(\psi)$ to the SAT solver, and if a satisfying assignment is found, add the missing information back in at the time we construct a model for $\phi$.

\subsection{Top-down Grounding over a Bound Structure}\label{subsec:GTD}

Algorithm \ref{alg:TD-Bound-Gnd} produces a grounding of $\phi$ over a bound structure 
$\hat{\cA}$ for \cA. $Gnd$ and $Simpl$ are defined by mutual recursion. $Gnd$ performs expansions and substitutions, while $Simpl$ performs lookups in $\hat{\given}$ to see if the grounding of a sub-formula may be left out. $Eval$ provides the base cases, evaluating ground atoms over $\instvocab \cup \expvocab \cup \tilde{A} \cup \omega$ in $\hat{\given}$.


\begin{algorithm}
\caption{Top-Down Grounding over Bound Structure $\hat{\given}$ for $\phi$ and \cA}
\label{alg:TD-Bound-Gnd}
\begin{equation*}
Gnd_{\hat{\given}}(\phi,\theta)=
\begin{cases}
Eval_{\hat{\given}}(P, \theta)                                    & \phi \textrm{ is an atom } P(\bar{x})\\
\neg Eval_{\hat{\given}}(P, \theta)                               & \phi \textrm{ is a negated atom }  \neg P(\bar{x}) \\
\bigwedge_i Simpl_{\hat{\given}}(\psi_{i}, \theta) & \phi=\bigwedge_i \psi_i \\
\bigvee_i Simpl_{\hat{\given}}(\psi_{i}, \theta)   & \phi=\bigvee_i \psi_i \\
\bigwedge_{a\in \givendom} Simpl_{\hat{\given}}(\psi,\theta \cup (x\slash\tilde{a}))           & \phi=\forall x\ \psi \\
\bigvee_{a\in \givendom} Simpl_{\hat{\given}}(\psi,\theta \cup (x\slash\tilde{a}))            & \phi=\exists x\ \psi
\end{cases}
\end{equation*}

\begin{equation*}\label{eq:evalInst}
Eval_{\hat{\given}}(P, \theta) = 
\begin{cases}
\top				& \hat{\given} \models P[\theta]\\
\bot				& \hat{\given} \models \neg P[\theta]\\
P(\bar{x})[\theta]	& \mathrm{o.w}
\end{cases}
\end{equation*}

\begin{equation*}
Simpl_{\hat{\given}}(\psi, \theta)  = 
\begin{cases}
\top				& \hat{\given} \models \subfor{\psi}[\theta]\\
\bot				& \hat{\given} \models \neg \subfor{\psi}[\theta]\\
Gnd_{\hat{\given}}(\psi,\theta)		& \mathrm{o.w}
\end{cases}
\end{equation*}
\end{algorithm}


The stronger $\hat{\given}$ is, the smaller the ground formula produced by Algorithm \ref{alg:TD-Bound-Gnd}. If we set $\hat{\given}$ to be undefined everywhere (i.e., to just give the domain), then Algorithm \ref{alg:TD-Bound-Gnd} produces $\mathrm{NaiveGnd}_{\cA}(\phi,\emptyset)$. If $\hat{\given}$ is set to $\given$, we get the reduced grounding obtained by evaluating instance symbols out of $\mathrm{NaiveGnd}_{\cA}(\phi)$. 

\begin{proposition}
Algorithm~\ref{alg:TD-Bound-Gnd} produces a grounding of $\phi$ over $\hat{\cA}$.
\end{proposition}

\subsection{Autarkies and Autark Subformulas}

In the literature, an \emph{autarky} \cite{Monien:1985:SSL:1703256.1703272} is informally a ``self-sufficient`` model for some clauses which does not affect the remaining clauses of the formula. An autark subformula is a subformula which is satisfied by an autarky. To see how an autark subformula may be produced during grounding, let $\lambda = \gamma_{1} \vee \gamma_{2}$ and imagine that the value of subformula $\gamma_{1}$ is true according to our bound structure. Then $\lambda$ will be true, regardless of the value of $\gamma_{2}$, and the grounder will replace its subformula with its truth value, whereas in the case of naive grounding, the grounder does not have that information during the grounding. So it generates the set of clauses for this subformula as: $\{(\neg \lambda \vee \gamma_{1} \vee \gamma_{2}), (\neg \gamma_{1} \vee \lambda), (\neg \gamma_{2} \vee \lambda)\}$. Now the propagation of the truth value of $\lambda_{1}$ and subsequently $\lambda$, results in elimination of all the three clauses, but the set of clauses generated for $\gamma_{2}$ will remain in the CNF formula. We call $\gamma_{2}$ and the clauses made from that subformula autarkies. 

The example suggests that this is a common phenomena and that the number of autarkies might be quite large in many groundings, as will be seen in Section \ref{sec:exp}.

\section{Lifted Unit Propagation Structures}\label{sec:LUP}

In this section we define $\mathcal{LUP}(\phi, \given)$, and 
a method for constructing it. 

\begin{definition}[$\mathcal{LUP}(\phi, \given)$]
Let $\mathrm{Units}$ denote the set of unit clauses that appears during the execution 
of UP on $\mathrm{CNF(NaiveGnd_{\cA}(\phi))}$.
The LUP structure for $\phi$ and \cA is the unique bound structure for $\phi$ and \cA
for which: \begin{equation}
\mathlarger{\chi}_{\ceil{\psi}}^{\cA}(\bar{a})  = 
\begin{cases}
\top &  \ceil{\psi}(\tilde{\bar{a}}) \in \mathrm{Units} \\           
\bot & \neg \ceil{\psi}(\tilde{\bar{a}}) \in \mathrm{Units} \\  
\infty  & \mathrm{o.w}
\end{cases}
\end{equation}
\end{definition}

Since Algorithm~\ref{alg:TD-Bound-Gnd} produces a grounding, according 
to Definition~\ref{def:gndpar}, for any bound structure, it produces a 
grounding for $\phi$ over $\mathcal{LUP}(\phi, \given)$.

To construct $\mathcal{LUP}(\phi, \given)$, we use an inductive definition obtained from $\phi$. In this inductive definition, we use distinct vocabulary symbols for the sets of tuples which $\hat{\given}$ sets to true and false. The algorithm works based on the notion of \emph{True (False) bounds}:

\begin{definition}[Formula-Bound]\label{def:TFBound} A \emph{True (resp. False) bound} for a subformula $\psi(\bar{x})$ according to bound structure $\hat{\given}$ is the relation denoted by $T_{\psi}$ (resp. $F_{\psi}$) such that:
\begin{enumerate}
\item $\bar{a} \in T_{\psi} \Leftrightarrow \subfor{\psi}^{\hat{\given}}(\bar{a}) = \top$
\item $\bar{a} \in F_{\psi} \Leftrightarrow \subfor{\psi}^{\hat{\given}}(\bar{a}) = \bot$
\end{enumerate}
\end{definition}

Naturally, when $\subfor{\psi}^{\hat{\given}}(\bar{a}) = \infty$, $\bar{a}$ is not contained in either $T_{\psi}$ or $F_{\psi}$.

The rules of the inductive definition are given in Table \ref{tab:propagation-rules}. 
These rules rules may be read as rules of FO(ID), the extension of classical logic with 
inductive definitions under the well-founded semantics \cite{VanGelder:1991:WSG:116825.116838,Denecker:2008:LNI:1342991.1342998}, with free variables implicitly universally quantified.
The \emph{type} column indicates the type of the subformula, and the \emph{rules} columns identify the rule for this subformula. Given a $\instvocab$-structure $\given$, we may evaluate the definitions on $\given$, thus obtaining a set of concrete bounds for the subformulas of $\phi$. The rules reflect the reasoning that UP can do. For example consider rule $(\vee_{i}\psi_{i})$ of $\dt$ for $\gamma(\bar{x})=\psi_1(\bar{x}_1)\vee\dots\vee \psi_N (\bar{x}_N)$, and for some $i\in\{1,\dots,N\}$:

\[ T_{\psi_i}(\bar{x}_i) \rul T_\gamma(\bar{x}) \wedge \bigwedge_{j\neq i} F_{\psi_j}(\bar{x}_j).\] 
This states that when a tuple $\bar{a}$ satisfies $\gamma$ but falsifies all disjuncts, 
$\psi_j$, of $\gamma$ except for one, namely $\psi_i$, then it must satisfy $\psi_i$. As a starting point, we know the value of the instance predicates, and we also assume that $\phi$ is $\given$-satisfiable.

\begin{center}
\begin{table}[bt]
\begin{center}
\begin{tabular}{c}
\begin{tabular}{| l | r l l |}

\hline
\multicolumn{1}{|c|}{type} & \multicolumn{3}{|c|}{$\dt$ rules} \\
\hline
$(\vee_{i}\psi_{i})$   & $\ T_{\psi_i}(\bar{x}_i)$ & $\leftarrow$ & $ T_\gamma(\bar{x}) \wedge \bigwedge_{j\neq i} F_{\psi_j}(\bar{x}_j)$, for each $i$\\
$(\wedge_{i}\psi_{i})$ & $\ T_{\psi_i}(\bar{x}_i)$ & $\leftarrow$ & $T_\gamma(\bar{x})$, for each $i$ \\
$\exists y \;\psi(\bar{x}, y)$   & $T_\psi(\bar{x},y)$ & $\leftarrow$ & $ T_\gamma(\bar{x}) \wedge \forall y'\!\neq \!y \;\;F_\psi(\bar{x},y')$ \\
$\forall y \;\psi(\bar{x}, y)$   & $T_\psi(\bar{x},y)$   & $\leftarrow$ & $ T_\gamma(\bar{x})$ \\
$P(\bar{x})$ & $\ T_P(\bar{x})$ & $\leftarrow$ & $T_\gamma(\bar{x})$ \\
$\neg P(\bar{x})$ & $\ F_P(\bar{x})$ & $\leftarrow$ & $T_\gamma(\bar{x})$ \\
\hline

\end{tabular}

\hspace{1.5mm}

\begin{tabular}{| l | r l l |}
\hline
\multicolumn{1}{|c|}{type} & \multicolumn{3}{|c|}{$\ut$ rules} \\
\hline
$(\vee_{i}\psi_{i})$ & $\ T_\gamma(\bar{x})$ & $\leftarrow$ & $\bigvee_{i}T_{\psi_i}(\bar{x}_i)$, for each $i$\;\;\;\;\;\;\;\;\;\\
$(\wedge_{i}\psi_{i})$  & $\ T_\gamma(\bar{x})$ & $\leftarrow$ & $\bigwedge_{i}T_{\psi_i}(\bar{x}_i)$, for each $i$ \\
$\exists y \;\psi(\bar{x}, y)$ & $\ T_\gamma(\bar{x})$ & $\leftarrow$ & $\exists y\ T_\psi(\bar{x},y)$ \\
$\forall y \;\psi(\bar{x}, y)$ & $\ T_\gamma(\bar{x})$ & $\leftarrow$ & $\forall y\ T_\psi(\bar{x},y)$ \\
$P(\bar{x})$ & $\ T_\gamma(\bar{x})$ & $\leftarrow$ & $ T_P(\bar{x})$ \\
$\neg P(\bar{x})$  & $\ T_\gamma(\bar{x})$ & $\leftarrow$ & $F_P(\bar{x})$ \\

\hline
\end{tabular}
\\

\\
\begin{tabular}{| l | r l l |}
\hline
\multicolumn{1}{|c|}{type} & \multicolumn{3}{|c|}{$\df$ rules} \\
\hline
$(\vee_{i}\psi_{i})$   & $\ F_{\psi_i}(\bar{x_i})$ & $\leftarrow$ & $F_\gamma(\bar{x})$, for each $i$ \\
$(\wedge_{i}\psi_{i})$ & $\ F_{\psi_i}(\bar{x}_i)$ & $\leftarrow$ & $F_\gamma(\bar{x}) \wedge \bigwedge_{ j\neq i} T_{\psi_j}(\bar{x}_j)$, for each $i$\\
$\exists y \;\psi(\bar{x}, y)$   & $F_\psi(\bar{x},y)$   & $\leftarrow$ & $F_\gamma(\bar{x})$ \\
$\forall y \;\psi(\bar{x}, y)$   & $F_\psi(\bar{x},y)$   & $\leftarrow$ & $F_\gamma(\bar{x}) \wedge \forall y'\!\neq\! y \;\;T_\psi(\bar{x},y')$ \\
$P(\bar{x})$         & $\ F_P(\bar{x})$ & $\leftarrow$ & $F_\gamma(\bar{x})$ \\
$\neg P(\bar{x})$    & $\ T_P(\bar{x})$ & $\leftarrow$ & $F_\gamma(\bar{x})$ \\
\hline
\end{tabular}

\hspace{1.5mm}

\begin{tabular}{| l | r l l |}
\hline
\multicolumn{1}{|c|}{type} & \multicolumn{3}{|c|}{$\uf$ rules} \\
\hline
$(\vee_{i}\psi_{i})$    & $\ F_\gamma(\bar{x})$ & $\leftarrow$ & $\bigwedge_{i}F_{\psi_i}(\bar{x}_i)$, for each $i$\;\;\;\;\;\;\;\;\\
$(\wedge_{i}\psi_{i})$  & $\ F_\gamma(\bar{x})$ & $\leftarrow$ & $\bigvee_{i}F_{\psi_i}(\bar{x}_i)$, for each $i$\\
$\exists y \;\psi(\bar{x}, y)$ & $\ F_\gamma(\bar{x})$ & $\leftarrow$ & $\forall y\ F_\psi(\bar{x},y)$ \\
$\forall y \;\psi(\bar{x}, y)$ & $\ F_\gamma(\bar{x})$ & $\leftarrow$ & $\exists y\ F_\psi(\bar{x},y)$ \\
$P(\bar{x})$ & $\ F_\gamma(\bar{x})$ & $\leftarrow$ & $F_P(\bar{x})$ \\
$\neg P(\bar{x})$  & $\ F_\gamma(\bar{x})$ & $\leftarrow$ & $T_P(\bar{x})$ \\
\hline
\end{tabular} 

\\
\vspace{0.55mm}

\end{tabular}
\end{center}
\caption{Rules for Bounds Computation}
\label{tab:propagation-rules}
\end{table}
\end{center}

\begin{example}\label{ex:bound}
Let $\phi=\forall x\ \neg I_1(x)\vee E_1(x)$, $\sigma=\{I_1,I_2\}$, and $\given=\big(\{1,2,3,4\}; I_1^\given=\{1\}\big)$. The relevant rules from Table (\ref{tab:propagation-rules}) are:
\begin{align*}
\ T_{\neg I_1(x)\vee E_1(x)}(x) & \leftarrow T_{\phi} \\
\ T_{I_1}(x)                    & \leftarrow I_1(x) \\
\ F_{\neg I_1(x)}(x)            & \leftarrow T_{I_1}(x) \\
\ T_{E_1(x)}(x)                 & \leftarrow T_{\neg I_1(x)\vee E_1(x)}(x) \wedge F_{\neg I_1(x)}(x) \\
\ T_{E_1}(x)                    & \leftarrow T_{E_1(x)}(x)
\end{align*}
\noindent We find that $T_{E_1}=\{1\}$; in other words: $E_1(1)$ is true in each model of $\phi$ expanding $\given$.
\end{example}

Note that this inductive definition is monotone, because $\phi$ is in \emph{Negation Normal Form (NNF)}.

\subsection{LUP Structure Computation}

 \begin{algorithm}[b]
\caption{Computation of {$\cal LUP$}($\phi, \given$)}
\label{alg:boundsComp}
\begin{algorithmic}[1]
\STATE Construct the rules $\{\ut, \dt, \uf\}$
\STATE Compute bounds by evaluating the inductive definition $\{\dt,\uf\}$
\IF {Bounds are inconsistent}
       \RETURN ``$\given$ has no solution''
\ENDIF
\STATE Throw away $T_{\psi}(\bar{x})$ for all non-atomic subformulas $\psi(\bar{x})$ \label{line:throw}
\STATE Compute new bounds by evaluating the inductive definition $\{\ut\}$ 
\RETURN LUP structure constructed from the computed bounds, according to Definition \ref{def:TFBound} .
\end{algorithmic}
\end{algorithm}

Our method for constructing ${\cal LUP}$($\phi, \given$) is given in Algorithm 3.
Several lines in the algorithm require explanation. 
In line 1, the $\df$ rules are omitted from the set of constructed rules. 
Because $\phi$ is in NNF, the $\df$ rules do not contribute any information 
to the set of bounds.  To see this, observe that every $\df$ rule has an 
atom of the form $F_{\gamma}(\bar{x})$ in its body.   Intuitively, for one of these 
rules to contribute a defined bound, certain information must have previously 
been obtained regarding bounds for its parent.   It can be shown, by induction, 
that, in every case, the information about a bound inferred by an application of a
$\df$ rule must have previously been inferred by a $\uf$ rule.
In line 2 of the algorithm we compute bounds using only the two sets of rules, $\dt$ and $\uf$. This is justified by the fact that applying $\{\ut, \dt, \uf\}$ to a fixpoint has the same effect as applying $\{\dt, \uf\}$ to a fixpoint and then applying the $\ut$ rules afterwards. So we postpone the execution of the $\ut$ rules to line 7.


Line 3 checks for the case that the definition has no model, which is to say 
that the rules allow us to derive that some atom is both in the true bound 
and the false bound for some subformula.   This happens exactly when UP 
applied to the naive grounding would detect inconsistency.

Finally, in lines 6 and 7 we throw away the true bounds for all non-atomic subformulas, 
and then compute new bounds by evaluating the $\ut$ rules, taking already computed 
bounds (with true bounds for non-atoms set to empty) as the initial bounds in the 
computation.   To see why,  observe that the true bounds computed in line 2 
are based on the assumption that $\phi$ is $\given$-satisfiable. 
So $\subfor{\phi}$ is set to true which stops the top-down bounded grounding algorithm 
of Section \ref{subsec:GTD} from producing a grounding for $\phi$. 
That is because the $Simpl$ function, considering the true bound for the $\phi$, simply 
returns $\top$ instead of calling $Gnd_{\hat{\given}}(.,.)$ on subformulas of the $\phi$. 
This also holds for all the formulas with true-bounds, calculated this way, except for the atomic 
formulas.  So, we delete these true bounds based on the initial unjustified assumption, and then 
construct the correct true bounds by application of the $\ut$ rules, in line 7. This is the main reason for postponing the execution of $\ut$ rules.

\section{Bottom-up Grounding over Bound Structures}\label{sc:bugnd}

The grounding algorithm we use in Enfragmo constructs a grounding by a bottom-up 
process that parallels database query evaluation, based on an extension of the 
relational algebra.  We give a rough sketch of the method here: further 
details can be found in, e.g., \cite{Mohebali:MSc-2006,PLTG:2007:IJCAI}. Given a structure 
(database) $\cA$, a boolean query is a formula $\phi$ over the vocabulary 
of $\cA$, and query answering is evaluating whether $\phi$ is true, i.e.,  $\cA \models \phi$.  
In the context of grounding, $\phi$ has some additional vocabulary beyond that of $\cA$, and
producing a reduced grounding involves evaluating out the instance vocabulary, and 
producing a ground formula representing the expansions 
of $\cA$ for which $\phi$ is true.

For each sub-formula $\alpha(\bar{x})$ with free variables
$\bar{x}$, we call the set of reduced groundings for $\alpha$ under all possible
ground instantiations of $\bar{x}$ an answer to $\alpha(\bar{x})$. We represent
answers with tables on which the extended algebra operates.
An X-relation, in databases, is a $k$-ary relation associated with a $k$-tuple of 
variables X, representing a set of instantiations of the variables of X.  Our grounding 
method uses extended X-relations, in which each tuple $\bar{a}$ is associated 
with a formula.   In particular, if $R$ is the answer to $\alpha(\bar{x})$, then 
$R$ consists of the pairs $(\bar{a},\alpha(\tilde{\bar{a}}))$. Since a sentence has no free variables, the answer to a sentence $\phi$ is a 
zero-ary extended X-relation, containing a single pair $( \langle\rangle, \psi )$, 
associating the empty tuple with formula $\psi$, which is a reduced grounding 
of $\phi$.

The relational algebra has operations corresponding to each connective and quantifier 
in FO: complement (negation); join (conjunction); union (disjunction), 
projection (existential quantification); division or quotient (universal quantification).  
Each generalizes to extended X-relations. If $(\bar{a}, \alpha(\tilde{\bar{a}})) \in \cR$ then we write $\delta_{\cR}(\bar{a}) = \alpha(\tilde{\bar{a}})$. For example, the join of extended $X$-relation $\cR$ and extended $Y$-relation $\cS$ 
(both over domain $A$), denoted 
$\cR \Join\cS$, is the extended $X \cup Y$-relation $\{(\bar{a}, \psi) \mid \bar{a}:X\cup
Y\rightarrow A,{\bar{a}} |_X\in \cR, {\bar{a}} |_Y\in \cS, \mbox{ and } \psi
= \delta_{\cR}({\bar{a}} |_X)\wedge \delta_{\cS}({\bar{a}} |_Y)\};$
It is easy to show that, if $\cR$ is an answer to $\alpha_1(\bar{x})$ and $\cS$ is an 
answer to $\alpha_2(\bar{y})$ (both wrt $\cA$), then 
$\cR \Join\cS$ is an answer to $\alpha_1(\bar{x}) \land \alpha_2(\bar{y})$. The analogous 
property holds for the other operators.

To ground with this algebra, we define the answer to atomic formula $P(\bar{x})$
as follows. If $P$ is an instance predicate, the answer is the set of tuples
$(\bar{a}, \top)$, for $\bar{a} \in P^{\cA}$. If $P$ is an expansion predicate,
the answer is the set of all tuples $(\bar{a}, P(\bar{a}))$, for $\bar{a}$ a
tuple of elements from the domain of $\cA$. Then we apply the algebra
inductively, bottom-up, on the structure of the formula.
At the top, we obtain the answer to $\phi$, which is a relation containing 
only the pair $(\langle\rangle, \psi )$, where $\psi$ is a 
reduced grounding of $\phi$ wrt $\cA$.

\begin{example}
Let $\sigma=\{P\}$ and $\varepsilon=\{E\}$, and let $\cA$ be a
$\sigma$-structure with $P^{\cA}=\{(1,2,3), (3,4,5)\}$.  The following extended
relation $\cR$ is an answer to $\phi_1 \equiv P(x,y,z) \wedge E(x,y) \wedge E(y,z)$:

\begin{small}
\begin{center}
\begin{tabular}{|c|c|c|c|}
\hline
$x$ & $y$ & $z$ & $\psi$\\
\hline
1 & 2 & 3 & $E(1,2)\wedge E(2,3)$\\
\hline
3 & 4 & 5 & $E(3,4)\wedge E(4,5)$\\
\hline
\end{tabular}
\end{center}
\end{small}

\noindent Observe that $\delta_{\cR}(1,2,3) = E(1,2)\wedge E(2,3)$
is a reduced grounding of
$\phi_1[(1,2,3)]=P(1,2,3) \wedge E(1,2) \wedge E(2,3)$, and
$\delta_{\cR}(1,1,1) = \bot$ is a reduced grounding of $\phi_1[(1,1,1)]$.

The
following extended relation is an answer to $\phi_2 \equiv \exists z \phi_1$:

\begin{small}
\begin{center}
\begin{tabular}{|c|c|c|}
\hline
$x$ & $y$ & $\psi$\\
\hline
1 & 2 & $E(1,2) \wedge E(2,3)$\\
\hline
3 & 4 & $E(3,4) \wedge E(4,5)$\\
\hline
\end{tabular}
\end{center}
\end{small}

\noindent Here, $E(1,2) \wedge E(2,3)$ is a reduced grounding of
$\phi_2[(1,2)]$. Finally, the following represents an answer to
$\phi_3 \equiv \exists x \exists y \phi_2$, where the single formula is a
reduced grounding of $\phi_3$.

\begin{small}
\begin{center}
\begin{tabular}{|c|}
\hline
$\psi$\\
\hline
$[E(1,2) \wedge E(2,3)] \vee [E(3,4) \wedge E(4,5)]$ \\
\hline
\end{tabular}
\end{center}
\end{small}

\end{example}

To modify the algorithm to ground using ${\cal LUP}$($\phi, \given$) we need only change the base case for expansion predicates. To be precise, if P is an expansion predicate we set the answer to $P(\bar{x})$ to the set of pairs $( \bar{a}, \psi )$ such that:
\[
\psi = 
\begin{cases}
P(\tilde{\bar{a}}) & \mathrm{if} \;\; P^{{\cal LUP}(\phi, \given)}(\bar{a}) = \infty \\           
\top &  \mathrm{if} \;\; P^{{\cal LUP}(\phi, \given)}(\bar{a}) = \top \\  
\bot  &  \mathrm{if} \;\; P^{{\cal LUP}(\phi, \given)}(\bar{a}) = \bot.
\end{cases}
\]

Observe that bottom-up grounding mimics the second phase of Algorithm \ref{alg:boundsComp}, i.e., a bottom-up truth propagation, except that it also propagates the falses. So, for bottom up grounding, we can omit line 7 from Algorithm \ref{alg:boundsComp}.

\begin{proposition}
Let $( \langle\rangle, \psi )$ be the answer to sentence $\phi$ wrt $\given$ after LUP initialization, then: 


\[
Gnd_{\mathcal{LUP}(\phi, \given)}(\phi,\emptyset) \equiv \psi
\]

\noindent where $Gnd_{\mathcal{LUP}(\phi, \given)}(\phi,\emptyset)$ is the result of top-down grounding Algorithm \ref{alg:TD-Bound-Gnd} of $\phi$ over LUP structure $\mathcal{LUP}(\phi, \given)$.
\end{proposition}

This bottom-up method uses only the reduct of $\mathcal{LUP}(\phi, \given)$ defined by $\instvocab \cup \expvocab \cup \tilde{A}$, not the entire LUP structure.

\section{Experimental Evaluation of LUP}\label{sec:exp}

In this section we present an empirical study of the effect of LUP 
on grounding size and on grounding and solving times.   We also compare 
LUP with GWB in terms of these same measures.   The implementation 
of LUP is within our bottom-up grounder Enfragmo, as described in this 
paper, and the implementation of GWB is in the top-down grounder \gidl, 
which is described in \cite{WittocxMD:08,DBLP:journals/jair/WittocxMD10}. \gidl has several parameters to control the precision of the bounds computation. In our experiments we use the default settings.
We used \textsc{MiniSat} as the ground solver for Enfragmo.   \gidl produces an output specifically for the ground 
solver \textsc{MiniSat(ID)}, and together they form the IDP system \cite{lash08/WittocxMD08}.

We report data for instances of three problems: Latin Square Completion, Bounded Spanning Tree and Sudoku.    The instances 
are latin\_square.17068* instances of Normal Latin Square Completion, the 104\_rand\_45\_250\_* and 104\_rand\_35\_250\_* instances of BST, and 
the ASP contest 2009 instances of Sudoku from the Asparagus repository\footnote{http://asparagus.cs.uni-potsdam.de}.   All experiments were run on a Dell Precision T3400 computer with a 
quad-core 2.66GHz Intel Core 2 processor having 4MB cache and 8GB of RAM, running CentOS 5.5 with Linux kernel 2.6.18.

In Tables 2 and 4, columns headed ``Literals'' or ``Clauses'' give the number of 
literals or clauses in the CNF formula produced by Enfragmo without LUP (our baseline), 
or these values for other grounding methods expressed as a percentage of the 
baseline value.   In Tables 3 and 5, all values are times seconds.
All values give are means for the entire collection of instances.   Variances are not 
given, because they are very small.    We split the instances of 
BST, into two sets, based on the number of nodes
(35 or 45), because these two groups exhibit somewhat different behaviour, but
within the groups variances are also small.
In all tables, the minimum (best) values for each row are in bold face type, to 
highlight the conditions which gave best performance.
 
Table 2 compares the sizes of CNF formulas produced by Enfragmo without LUP (the base line) with the formulas obtained by running UP on the baseline formulas and by running Enfragmo with LUP. Clearly LUP reduces the size at least as much as UP, and usually reduces the size 
much more, due to the removal of autarkies. 

\begin{table*}[htb]
\begin{center}
\begin{tabular}{|c|c|c|c|c|c|c|}
\hline
& \multicolumn{2}{|c|}{Enfragmo}& \multicolumn{2}{|c|}{Enfragmo+UP (\%)}& \multicolumn{2}{|c|}{Enfragmo+LUP (\%)}\\
\hline
Problem&Literals&Clauses&Literals&Clauses&Literals&Clauses\\
\hline
Latin Square&7452400&2514100&\textbf{0.07}&\textbf{0.07}&\textbf{0.07}&\textbf{0.07}\\
\hline
BST 45&22924989&9061818&0.96&0.96&\textbf{0.24}&\textbf{0.24}\\
\hline
BST 35&8662215&3415697&0.95&0.96&\textbf{0.37}&\textbf{0.37}\\
\hline
Sudoku&2875122&981668&0.17&0.18&\textbf{0.07}&\textbf{0.08}\\
\hline
\end{tabular}
\end{center}
\caption{Impact of LUP on the size of the grounding.  The first two columns give the 
numbers of literals and clauses in groundings produced by Enfragmo without LUP
(the baseline).
The other columns give these measures for formulas produced 
by executing UP on the baseline groundings (Enfragmo+UP), and 
for groundings produced by Enfragmo with LUP (Enfragmo+LUP),
expressed as a fraction baseline values.}
\vspace{0mm}
\label{tab:LUP-size}
\end{table*}

\begin{table*}[htb]
\begin{center}
\begin{tabular}{|c|c|c|c|c|c|c|c|c|c|}
\hline
&\multicolumn{3}{|c|}{Enfragmo}& \multicolumn{3}{|c|}{Enfragmo with LUP}& \multicolumn{3}{|c|}{Speed Up Factor}\\
\hline
Problem&Gnd&Solving&Total&Gnd&Solving&Total&Gnd&Solving&Total\\
\hline
Latin Square&\textbf{0.89}&1.39&\textbf{2.28}&3.27&\textbf{0.34}&3.61&-2.38&1.05&-1.33\\
\hline
BST 45&6.08&7.56&13.64&\textbf{2}&\textbf{1.74}&\textbf{3.74}&4.07&5.82&9.9\\
\hline
BST 35&2.13&2.14&4.27&\textbf{1.07}&\textbf{0.46}&\textbf{1.53}&1.06&1.68&2.74\\
\hline
Sudoku&\textbf{0.46}&1.12&\textbf{1.59}&2.08&\textbf{0.26}&2.34&-1.62&0.86&-0.76\\
\hline
\end{tabular}
\end{center}
\caption{Impact of LUP on reduction in both grounding and (SAT) solving time. Grounding time here includes LUP computations and CNF generation.}
\label{tab:LUP-time}
\end{table*}

Total time for solving a problem instance is composed of grounding time and SAT solving time. 
Table~\ref{tab:LUP-time} compares the grounding and SAT solving time with and without LUP bounds. It is evident that the SAT solving time is always reduced with LUP. This reduction is due to the elimination of the unit clauses and autark subformulas from the grounding. Autark subformula elimination also affects the time required to convert the ground formula to CNF which reduces the grounding time, but in some cases the overhead imposed by LUP computation may not be made up for by this reduction. As the table shows, when LUP outperforms the normal grounding we get factor of 3 speed-ups, whereas when it loses to normal grounding the slowdown is by a factor of 1.5. 

Table \ref{tab:LUP-GIDL-size}  compares the size reductions obtained by LUP and 
by GWB in  \gidl.  The output of \gidl contains clauses and rules. 
The rules are transformed to clauses in (\textsc{MiniSat(ID)}). 
The measures reported here are after that transformation.
 LUP reduces the size much more than GWB, in most of the cases. This stems from the fact that \gidl's bound computation does not aim for completeness wrt unit propagation. This also affects the solving time because the CNF formulas are much smaller with LUP as shown in Table~\ref{tab:LUP-GIDL-time}. 
  Table \ref{tab:LUP-GIDL-time} shows that Enfragmo with LUP and \textsc{MiniSat} is always faster than \gidl with \textsc{MiniSat(ID)} with or without bounds, and it is in some cases faster than Enfragmo without LUP.

\begin{table*}[b]
\begin{center}
\begin{tabular}{|c|c|c|c|c|c|c|c|c|}
\hline
&\multicolumn{2}{|c|}{Enfragmo (no LUP)}& \multicolumn{2}{|c|}{\gidl (no bounds)}& \multicolumn{2}{|c|}{Enfragmo with LUP}& \multicolumn{2}{|c|}{\gidl with bounds}\\
\hline
Problem&Literals&Clauses&Literals&Clauses&Literals&Clauses&Literals&Clauses\\
\hline
Latin Square&7452400&2514100&0.74&0.84&\textbf{0.07}&\textbf{0.07}&0.59&0.61\\
\hline
BST 45&22924989&9061818&0.99&1.02&\textbf{0.24}&\textbf{0.24}&0.25&\textbf{0.24}\\
\hline
BST 35&8662215&3415697&1.01&1.04&\textbf{0.37}&\textbf{0.37}&0.39&0.39\\
\hline
Sudoku&2875122&981668&0.56&0.6&\textbf{0.07}&\textbf{0.08}&0.38&0.39\\
\hline
\end{tabular}
\end{center}
\caption{Comparison between the effectiveness of LUP and \gidl Bounds on reduction in grounding size. The columns under Enfragmo show the actual grounding size whereas the other columns show the ratio of the grounding size relative to that of Enfragmo (without LUP).}
\label{tab:LUP-GIDL-size}
\end{table*}

\begin{table*}[htb]
\begin{small}
\begin{center}
\begin{tabular}{|c|c|c|c|c|c|c|c|c|c|c|c|c|}
\hline
&\multicolumn{3}{|c|}{Enfragmo}& \multicolumn{3}{|c|}{IDP}& \multicolumn{3}{|c|}{Enfragmo+LUP}& \multicolumn{3}{|c|}{IDP (Bounds)}\\
\hline
Problem&Gnd&Solving&Total&Gnd&Solving&Total&Gnd&Solving&Total&Gnd&Solving&Total\\
\hline
Latin Square&\textbf{0.89}&1.39&\textbf{2.28}&3&4.63&7.63&3.27&\textbf{0.34}&3.61&2.4&3.81&6.21\\
\hline
BST 45&6.08&7.56&13.64&7.25&20.84&28.09&2&\textbf{1.74}&\textbf{3.74}&\textbf{1.14}&4.45&5.59\\
\hline
BST 35&2.13&2.14&4.27&2.63&6.31&8.94&1.07&\textbf{0.46}&\textbf{1.53}&\textbf{0.67}&2.73&3.4\\
\hline
Sudoku&\textbf{0.46}&1.12&\textbf{1.59}&1.81&1.3&3.11&2.08&\textbf{0.26}&2.34&2.85&0.51&2.37\\
\hline
\end{tabular}
\end{center}
\end{small}
\caption{Comparison of solving time for Enfragmo and IDP, with and without LUP/bounds.}
\label{tab:LUP-GIDL-time}
\end{table*}


\section{Discussion}\label{sec:disc}

In the context of grounding-based problem solving, we have described a method
we call lifted unit propagation (LUP) 
for carrying out a process essentially equivalent to unit propagation before and 
during grounding.   Our experiments indicate that the method can substantially 
reduce grounding size -- even more than unit propagation itself, and sometimes 
reduce total solving time as well.     

Our work was motivated by the results of \cite{WittocxMD:08,DBLP:journals/jair/WittocxMD10}, which presented the method 
we have referred to as GWB.  In GWB, bounds on sub-formulas of the specification 
formula are computed without reference to an instance structure, and represented 
with FO formulas.  The grounding algorithm evaluates instantiations of these 
bound formulas on the instance structure to determine that 
certain parts of the naive grounding may be left out.   If the bound formulas 
exactly represent the information unit propagation can derive, then LUP 
and GWB are equivalent (though implemented differently).   However, generally 
the GWB bounds are weaker than the LUP bounds, for two reasons. 
First, they must be weaker, because no FO formula can define the bounds 
obtainable with respect to an arbitrary instance structure.  Second, to 
make the implementation in \gidl efficient, the computation of the bounds 
is heuristically truncated.   This led us to ask how much additional 
reduction in formula size might be obtained by the complete LUP method, 
and whether the LUP computation could be done fast enough for this 
extra reduction to be useful in practice.

Our experiments with the Enfragmo and \gidl grounders show that, at least 
for some kinds of problems and instances, using LUP can produce much smaller 
groundings than the GWB implementation in \gidl.    In our experiments, 
the total solving times for Enfragmo with ground solver \textsc{MiniSat} 
were always less than those of \gidl with ground solver \textsc{MiniSat(ID)}.   However, 
LUP reduced total solving time of Enfragmo with \textsc{MiniSat} significantly 
in some cases, and increased it --- albeit less significantly --- in others.   Since  there are many 
possible improvements of the LUP implementation, the question 
of whether LUP can be implemented efficiently enough to be used all the 
time remains unanswered.  

Investigating more efficient ways to do LUP, such 
as by using better data structures, is a subject for future work, as is 
consideration of other approximate methods such, as placing a heuristic 
time-out on the LUP structure computation, or dovetailing of the LUP
computation with grounding.
We also observed that the much of the reduction in grounding size obtained 
by LUP is due to identification of autark sub-formulas.   These cannot be 
eliminated from the naive grounding by unit propagation.  Further investigation 
of the importance of these in practice is another direction we are pursuing.
One more direction we are pursuing is the study of methods for deriving 
even stronger information that represented by the LUP structure, 
to further reduce ground formula size, and possibly grounding time as well.

\subsection*{Acknowledgements}
The authors are grateful to Marc Denecker, Johan Wittocx, Bill MacReady, 
Calvin Tang, Amir Aavani, Shahab Tasharoffi, Newman Wu, D-Wave Systems, 
MITACS, and NSERC.

\bibliographystyle{alpha}
\bibliography{mx} 
\end{document}

%% file: mx-def.tex
\newcommand{\ignore}[1]{} 

\newcommand{\gidl}{\textsc{GidL}\xspace}

\newcommand{\cA}{\ensuremath{{\cal{A}}}\xspace}

\newcommand{\cB}{\ensuremath{{\cal{B}}}\xspace}

\newcommand{\cR}{\ensuremath{{\cal{R}}}\xspace}
\newcommand{\cS}{\ensuremath{{\cal{S}}}\xspace}

\newcommand{\rul}{{}\leftarrow{}}

\newcommand{\model}{\cB}
\newcommand{\given}{\cA}
\newcommand{\givendom}{A}
\newcommand{\instvocab}{\sigma}
\newcommand{\expvocab}{\varepsilon}

\newcommand{\subfor}[1]{\lceil #1\rceil}

\newcommand{\ceil}[1]{\ensuremath{\lceil #1 \rceil}}

\newcommand{\muparrow}{\mathord{\uparrow}}
\newcommand{\mdownarrow}{\mathord{\downarrow}}

\newcommand{\df}{\mdownarrow f}
\newcommand{\uf}{\muparrow f}
\newcommand{\dt}{\mdownarrow t}
\newcommand{\ut}{\muparrow t}

\def\b0{\ensuremath{\bar{0}}}

\def\AC0{\ensuremath{\mathtt{AC^0}}}

\def\TC0{\ensuremath{\mathtt{TC^0}}}

\def\NC1{\ensuremath{\mathtt{NC^1}}}
\def\SAC1{\ensuremath{\mathtt{SAC^1}}}

\def\FAC0{\ensuremath{\mathtt{FAC^0}}}